\documentclass[letterpaper,10pt,final,balancelastpage,floats,aps,
twocolumn, a4paper]{revtex4}
\usepackage{amsmath}
\usepackage{amssymb}
\usepackage{longtable}
\usepackage{acronym}
\usepackage{graphicx}

\numberwithin{equation}{section}

\input{tcilatex}

\begin{document}

\title{Multireference Correlation in Long Molecules with
the Quadratic Scaling Density Matrix Renormalization Group}
\author{Johannes Hachmann, Wim Cardoen, and Garnet Kin-Lic Chan}
\affiliation{Department of Chemistry and Chemical Biology, Cornell University, Ithaca, NY
14853-1301, USA}
\date{\today }

\begin{abstract}
We have devised and implemented a local \textit{ab initio} Density Matrix
Renormalization Group (DMRG) algorithm to describe multireference correlations in large systems. For long molecules that are extended in one
of their spatial dimensions, we can obtain an exact
characterisation of correlation, in the given basis, with a cost that scales
only quadratically with the size of the system. The reduced scaling is
achieved solely through integral screening and without the construction of
correlation domains. We demonstrate the scaling, convergence, and robustness
of the algorithm in polyenes and hydrogen chains. We converge to exact
correlation energies (with 1-10$\mu $E$_{\text{h}}$ precision) in all cases
and correlate up to 100 electrons in 100 active orbitals. We further use our
algorithm to obtain exact energies for the metal-insulator transition in
hydrogen chains and compare and contrast our results with those from
conventional quantum chemical methods.
\end{abstract}

\maketitle

\section{Introduction}

The electronic structure of a chemical system features two types of electron
correlation. The first is \textit{nondynamic} correlation. This is
associated with the correlation of electrons in nearly degenerate valence
orbitals. The correct description of nondynamic correlation is necessary to
establish the qualitative features of chemical bonding. The second is 
\textit{dynamic} correlation. This is associated with excitations from
valence degrees of freedom into the many non-bonding orbitals. The multiple
weak excitations are responsible for establishing the detailed, quantitative
structure of the electronic wavefunction.

In general, a correct description of strong nondynamic correlation in large
systems is very difficult to obtain. When nondynamic correlation is
important (e.g. during bond breaking), a single determinant or electronic
configuration does not provide the correct qualitative structure of the
wavefunction. Instead, the delicate balance in the valence degrees of
freedom between the kinetic energy, which favours delocalisation, and the
Coulomb energy, which favours localisation, results in competing electronic
configurations and the correct electronic structure contains contributions
from multiple determinants with significant weights. Complete-Active-Space
Self-Consistent-Field (CASSCF) theories \cite{ROOS:1987:_casscf} correctly
describe this type of structure by expanding the wavefunction in the
complete space of the optimised valence (or \textquotedblleft
active\textquotedblright ) degrees of freedom, but do so at the cost of a
factorial scaling with the number of active electrons. Such calculations
with more than \textit{O}(10) electrons remain extremely difficult at this
time. Despite the impressive progress in local Generalised Valence Bond and
Coupled Cluster (CC) Theories (e.g. \cite%
{localgvb1,localgvb2,localgvb3,localgvb4,localgvb5,localgvb6,localgvb7,localgvb8,localcc1,localcc2,localcc3,localcc4,localcc5}%
) which provide some capacity to break e.g. single bonds, such approaches do
not possess the flexibility of a true multireference theory. The fundamental
challenge therefore remains to find a multireference electronic structure
method that is sufficiently flexible to correctly describe nondynamic
correlation, yet which exhibits a non-factorial scaling, and can thus be
applied to large systems.

In this work, we will adopt the more modest goal of answering the question
of how to describe nondynamic correlation in systems which are large in only
one out of their three spatial extents. In quasi-one-dimensional systems the
physics that is familiar from three-dimensions, is notably modified. This
is illustrated by the organic electronic materials (e.g. conjugated organic
polymers and carbon nanotubes) which exhibit unusual interacting electron
effects, arising from coupled quasi-one-dimensional motions of many electrons along the
conjugated $\pi $-backbone. As a simple example, in linear polyenes,
electron-electron interactions contribute to make the lowest excited singlet
state (the $^{1}2A_{g}$ state) one of double-excitation nature, rather
than the singly excited HOMO $\rightarrow $ LUMO state as one would expect
in a single particle picture \cite{polyene1, polyene2, polyene3}. A more
extreme example of this occurs in metallic nanotubes at low temperatures,
where at low energies there are no singly excited states; all low-energy
excitations are of collective nature, and the qualitative electronic
structure is of Luttinger liquid form \cite{nanotube1, nanotube2, nanotube3}.

Here we demonstrate that the Density Matrix Renormalization Group (DMRG),
which we and others have recently been investigating in quantum chemistry 
\cite{Shuai, Shuai2, Fano_PPP1, Fano_PPP2, MFOLP:2001:_dmrg,
MLPF:2003:_dmrg, WHI-MAR:1998:_dmrg, dmrg-ijqc, CHAN-HEA:2002:_dmrg,
CHAN-HEAD:2003:_dmrg, CHAN:2004:_dmrg, CHAN:2004b:_dmrg, dmrg-nh,
LEG-ROD:2003:_dmrg1, LEG-ROD:2003:_dmrg2, reiher, dmrg-rel, reiher2,
Ma_PPP1, Ma_PPP2, Ma_PPP3}, provides a solution to the question of how
to flexibly and efficiently describe nondynamic correlation in systems that
are large in one of their three spatial dimensions. Our analysis shows that
the DMRG behaves as a local, multireference, size-consistent/size-extensive,
and variational theory. From the intrinsic locality of the DMRG ansatz we
formulate a DMRG algorithm, denoted for convenience as LDMRG, that scales
only \textit{quadratically} with the size of the system, without any need
for an artifical imposition of orbital domains. The multireference nature of
the ansatz also eliminates any need for separately localised occupied and
virtual orbitals. Using this algorithm, we carry out numerically exact DMRG
calculations for long molecules, including polyenes in the $\pi $-active
space and metallic and insulating hydrogen chains where we correlate up to
100 active electrons in 100 active orbitals.

The structure of our discussion is as follows. We first introduce the DMRG
wavefunction ansatz in section \ref{sec:dmrgansatz}. There we discuss its
multireference, size-consistent and size-extensive, variational, and local
properties, and the implications for the design of a local DMRG algorithm.
In section \ref{sec:quadraticscaling} we show how a simple screening of
integral amplitudes results in a robust and naturally quadratic-scaling DMRG
algorithm. In section \ref{sec:calculations} we present calculations on
hydrogen molecular chains and polyenes in the $\pi $-active space and
demonstrate the size-extensivity, computational scaling, and convergence of
the LDMRG algorithm. As a difficult test of nondynamic correlation, we
further carry out calculations on the metal-insulator transition in hydrogen
chains for both symmetric and asymmetric bond stretching, and compare our
results against existing quantum chemical methods (section \ref{sec:mi}).
Finally, our conclusions are presented in section \ref{sec:conclusions}.

\section{The DMRG Ansatz}

\label{sec:dmrgansatz}

\subsection{DMRG and Matrix Product States}

\label{sec:mps} In previous work \cite{CHAN-HEA:2002:_dmrg}, we have
described the Renormalization Group formulation of the DMRG algorithm.
However, as related by \"{O}stlund and Rommer \cite{OSTLUND:1995:_mps,
OSTLUND:1997:_mps}, and subsequently developed by other authors (see e.g. 
\cite{VERSTRAETE:2004:_mps,SCHOLLWOCK:2005:_dmrg}), the DMRG is also
fruitfully analysed from the viewpoint of the underlying wavefunction
ansatz, the Matrix Product State (MPS). Such a formulation will be
convenient for our present discussion and we will recall the main points
below; for a full presentation, we refer the reader to the excellent review
by Schollw\"{o}ck \cite{SCHOLLWOCK:2005:_dmrg}.

Consider an $N$-particle system in a state $|\Psi \rangle $ spanned by $k$
orbitals. In occupation number representation, $|\Psi \rangle $ can be
expanded as 
\begin{equation}
|\Psi \rangle =\sum_{n_{1}n_{2}\ldots n_{k}}\psi _{n_{1}n_{2}\ldots
n_{k}}|n_{1}n_{2}\ldots n_{k}\rangle 
\end{equation}%
where $n_{i}=0,1;\ \sum_{i}n_{i}=N$. We now decompose the high dimensional
coefficient tensor $\psi $ into a chained matrix product via repeated
singular value decompositions (SVD). For example, if there are only two
orbitals, a singular value decomposition yields 
\begin{equation}
\psi _{n_{1}n_{2}}=\sum_{i}R_{i}^{n_{1}}\sigma _{i}R_{i}^{n_{2}}
\end{equation}%
where $\mathbf{R}^{n_{1}}$ and $\mathbf{R}^{n_{2}}$ are the singular vectors
and $\mathbf{\sigma }$ are the singular values. Similarly, for three
orbitals, $\psi _{n_{1}n_{2}n_{3}}$ can be decomposed via two singular value
decompositions as 
\begin{eqnarray}
\psi _{n_{1}n_{2}n_{3}} &=&\sum_{i}R_{i}^{n_{1}}\sigma _{i}S_{i}^{n_{2}n_{3}}
\notag \\
&=&\sum_{ij}R_{i}^{n_{1}}\sigma _{i}^{\prime }R_{ij}^{n_{2}}R_{j}^{n_{3}}
\end{eqnarray}%
where in the SVD of $\mathbf{S}_{i}$, all singular values have modulus one,
since $\mathbf{S}_{i}$ is an orthogonal matrix. In this way, through
repeated SVDs, the $k$-dimensional coefficient tensor can be decomposed as a
chain of matrix products, 
\begin{equation}
\psi _{n_{1}n_{2}\ldots n_{k}}=\text{Tr}\left\{ \mathbf{R}^{n_{1}}\mathbf{R}%
^{n_{2}}\ldots \mathbf{\sigma }\ldots \mathbf{R}^{n_{k}}\right\} .
\end{equation}%
So far the decomposition is exact since the $\mathbf{R}$ matrices are full
rank and will grow increasing large as the number of orbitals grows.

The \textit{Matrix Product State} which underlies the DMRG algorithm, arises
by truncating the maximum dimension of the $\mathbf{R}$ matrices to be at most $M\times M$%
, and thus with this restriction, we write the MPS as 
\begin{equation}
|\Psi \rangle =\sum_{n_{1}n_{2}\ldots n_{k}}\mathrm{Tr}\left\{ \mathbf{R}%
^{n_{1}}\mathbf{R}^{n_{2}}\ldots \mathbf{\sigma }\ldots \mathbf{R}%
^{n_{k}}\right\} |n_{1}n_{2}\ldots n_{k}\rangle .  \label{mps}
\end{equation}

We now establish the relationship between the MPS and the usual formulation
of the DMRG algorithm. Recall that any point in a DMRG sweep, the orbitals
are partitioned into two blocks: a left block (spanning orbitals $1,\ldots
,f $ say) and a right block (spanning orbitals $g=f+1,\ldots ,k$). Through
successive renormalisation transformations we obtain an adaptive many-body
basis of dimension $M$ to span the orbitals $1,\ldots ,f$; let us denote
these many-body states by $|l_{f}\rangle $. First we enlarge the left block
by adding the next orbital to give a super-block with an associated space $%
\{|l_{f}\rangle \}\otimes \{|n_{g}\rangle \}$. Next we renormalise this
space to form a new many-body basis $\{|l_{g}\rangle \}$ for the enlarged
block spanning orbitals $1,\ldots ,g$, as 
\begin{equation}
|l_{g}\rangle =\sum_{fn_{g}}R_{gf}^{n_{g}}|l_{f}n_{g}\rangle
\end{equation}%
where the rows of the matrix $\mathbf{R}^{n_{g}}$ are the $M$ eigenvectors
of the density matrix of the superblock $1,\ldots ,g$. After successive
renormalisations, we see that the renormalised states take on a matrix
product form, e.g. 
\begin{eqnarray}
|l_{h}\rangle &=&\sum_{gn_{h}}R_{hg}^{n_{h}}|l_{g}n_{h}\rangle  \notag \\
&=&\sum_{fgn_{g}n_{h}}R_{hg}^{n_{h}}R_{gf}^{n_{g}}|l_{f}n_{g}n_{h}\rangle 
\notag \\
&=&\ldots  \label{basisexpansion}
\end{eqnarray}%
where each $\mathbf{R}^{n_{i}}$ matrix is truncated to have maximum
dimension $M\times M$.

To complete the identification of the underlying DMRG wavefunction with the
Matrix Product State, we introduce the corresponding renormalised many-body
states $|r_{g}\rangle $ which span the orbitals $g=f+1,\ldots ,k$. In the
tensor-product space of the left and right blocks, we can write the full
wavefunction in the form 
\begin{equation}
|\Psi \rangle =\sum_{l_{f}r_{g}}\psi _{l_{f}r_{g}}|l_{f}r_{g}\rangle .
\end{equation}%
Performing an SVD, we obtain 
\begin{equation}
|\Psi \rangle =\sum_{f}|\bar{l}_{f}\rangle \sigma _{f}|\bar{r}_{f}\rangle .
\label{dmrgwf}
\end{equation}%
Substituting in the matrix product decomposition of the DMRG many-body basis
for the left and right block basis states from (\ref{basisexpansion}) in
eqn. (\ref{dmrgwf}) we identify the DMRG wavefunction with the Matrix
Product State (\ref{mps}). Consequently, the DMRG can be viewed as a
self-consistent optimisation algorithm for the Matrix Product State where
the renormalisation matrices $\mathbf{R}^{n_{i}}$ which parametrise the
ansatz are determined one by one from the density matrices of the blocks
after each blocking step in a DMRG sweep. The number of retained states in
the DMRG $M$ thus coincides with the dimensionality of the matrices that
parametrise the MPS. We note that the position of $\mathbf{\sigma }$ in the
Matrix Product State corresponds to the point of division between left and
right blocks in the DMRG algorithm. In principle, the DMRG wavefunction
varies with different block partitionings along a sweep, but in practice,
the variation is quite small.

\subsection{DMRG as a local, multireference, variational, size-consistent
ansatz for long systems}

Starting from the perspective above, let us summarise some features of the
DMRG/MPS ansatz.

\begin{enumerate}
\item {Variational}: Since we can associate a wavefunction with any DMRG
block configuration, and a DMRG energy is evaluated as an expectation value
of such a wavefunction, the energies appearing in the DMRG procedure are
strictly variational.

\item {Multireference}: It is clear that the Hartree-Fock reference has no
special significance in the DMRG state, and in particular, we do not order
or rank excitations relative to a single reference state. Furthermore, in
contrast to selected Configuration Interaction (CI) theories, none of the
coefficients of expansion $\psi_{n_1 n_2 \ldots n_k}$ are restricted to be
zero.

\item {Size-consistency}: Within a physical ordering of the orbitals on the
DMRG lattice, the Matrix Product State for two widely separated systems
factorises into the product of Matrix Product States for each system
separately. To see this, first arrange the orbitals into left and right
blocks, with the left block containing orbitals of the first system, and the
right block containing orbitals of the second system. Since there is no
coupling, the Matrix Product State for the total block configuration is a
product $|\Psi \rangle =|l\rangle |r\rangle $, where $|l\rangle $ is a
Matrix Product State for the first system considered alone (without changing
the orbital ordering) and similarly for $|r\rangle $. Consequently, the DMRG
energy is size-consistent.

\item {Locality and compactness}: The number of variational parameters in
the Matrix Product State is $O(M^{2}k)$ and its correlation length is
determined by $M$. Thus in any system with a finite quantum (i.e.
off-diagonal) correlation length along the DMRG lattice, we can obtain a
given accuracy in the energy per unit site with \textit{constant} $M$,
independent of the size of the system. In such cases, for a given accuracy,
the number of variational parameters in the DMRG scales linearly with the
size of the molecule. The restriction to given $M$ determines the finite
correlation length that is captured by the ansatz; there is no need to 
\textit{a priori} impose any orbital domains. Thus the DMRG is a naturally
local scaling ansatz, and so long as the determination of the energy is also
performed with an account of locality (e.g. through screening or multipole
expansion) a low-order scaling correlation theory arises. Indeed this is the
basis of the quadratic scaling algorithm in the next section. Note that a
finite correlation length implies only that we are away from a quantum
critical point; such wavefunctions need not be close to the Hartree-Fock
reference in any sense, as is indeed the case for systems with strong
interactions. Thus, the local correlation nature of the DMRG is different
from that of other local correlation methods (such as local CCSD) since
these require the \textit{correction} to the mean-field reference to be
small and to possess finite correlation length.

The DMRG ansatz possesses a further technical advantage. Since no
localisable Hartree-Fock reference is required, the favorable scaling of the
DMRG is obtained in any local basis, and does not in particular need
separate localisation in the occupied and virtual spaces. This is
particularly advantageous when modelling correlated states which possess a
shorter quantum correlation length than their parent mean-field reference
(e.g. systems with small Hartree-Fock bandgaps), for which orbital
localisation may be more difficult.

\item {Long molecules}: The Matrix Product State embodies a finite
correlation length as measured along the DMRG orbital lattice, rather than
as necessarily exists in the physical space. Consequently, we can only hold $%
M$ constant as the system size increases - and obtain the same relative
accuracy - if the locality in the physical system maps geometrically onto a
one-dimensional lattice, i.e. the system is extended in only one of its
three dimensions as in a long molecule is avoided. If this is not the case, we will require $M$ to
scale exponentially in the width of the system to maintain accuracy, much
like Full Configuration Interaction (FCI) or CASSCF theory. In practice, we
have shown that with reasonable $M$ we can still obtain highly accurate DMRG
energies even in non-one-dimensional molecular systems with up to $O(40)$
active orbitals - i.e. too large to treat using FCI theory - but to model
much larger extended non-one-dimensional networks of strongly interacting
electrons, further progress in the DMRG method will be required.
\end{enumerate}

\section{A quadratic scaling parallelised DMRG algorithm}

\label{sec:quadraticscaling}

The full computational scaling of a single conventional DMRG sweep is $%
O(M^{2}k^{4})+O(M^{3}k^{3})$. Here, the $O(k^{4})$ scaling arises in essence
from the number of two electron integrals $v_{ijkl}$ in the Hamiltonian $H$, written in
second quantisation as 
\begin{equation}
H=\sum_{ij}t_{ij}a_{i}^{\dag }a_{j}+\sum_{ijkl}v_{ijkl}a_{i}^{\dag
}a_{j}^{\dag }a_{k}a_{l}.  \label{heqn}
\end{equation}%

Recall that $M$ can be kept fixed, independent of system size in a long molecule. Thus to implement a quadratic scaling DMRG algorithm we need only screen the contributions from the 2 electron integrals. This can be achieved by working in a localized basis. 
(Note we can use any localised
orthonormal basis and we do not need to separately localise the occupied and
virtual spaces as is commonly required in local correlation methods. For
example, later in this work, we shall use the basis of overlap symmetrically
orthonormalised atomic orbitals). As is well understood, in a large system
described in a localised basis, the number of significant two-electron
integrals below a given threshold scales only quadratically as non-classical
Coulomb integrals i.e. integrals of the form $v_{ijkl}=\frac{1}{2}%
(i(1)l(1)|j(2)k(2))$ where $i(1),l(1)$ or $j(2),k(2)$ functions are widely
separated, vanish exponentially with the separation between $i,l$ or $j,k$
centres.

In the DMRG, we work with a number of intermediate combinations of operators
on each of the blocks of orbitals which are subsequently combined to
construct the full $H$ \cite{CHAN-HEA:2002:_dmrg, CHAN:2004:_dmrg}. A DMRG
sweep, consisting of $O(k)$ sweep iterations (each comprising a different
block configuration), requires $O(M^{2}k^{4})+O(M^{3}k^{3})$ time, $%
O(M^{2}k^{2})$ memory, and $O(M^{2}k^{3})$ disk storage. These asymptotic
costs originate from manipulating the two-index intermediate operators $%
A_{ij},B_{ij},P_{ij},Q_{ij}$ on the various blocks: 
\begin{align}
A_{ij(\in \text{blk})}^{\text{blk}}& = & & a_{i}a_{j}  \label{compop1} \\
B_{ij(\in \text{blk})}^{\text{blk}}& = & & a_{i}^{\dag }a_{j}
\label{compop2} \\
P_{ij(\notin \text{blk})}^{\text{blk}}& = & & \sum_{kl\in \text{blk}%
}v_{ijkl}a_{k}a_{l}  \label{compop3} \\
Q_{ij(\notin \text{blk})}^{\text{blk}}& = & & \sum_{kl\in \text{blk}%
}x_{ijkl}a_{k}^{\dag }a_{l}  \label{compop4} \\
x_{ijkl}& = & & v_{ijkl}-v_{jikl}-v_{ijlk}+v_{jilk}.
\end{align}

\begin{figure}[tbp]
\caption{Standard block configuration in DMRG. From left to right, $L$, $%
\circ _{L}$, $\circ _{R}$, $R$. }\centering{%
\includegraphics[width=7.5cm]{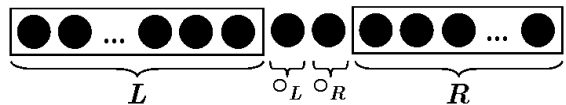}}
\label{fig:blockdiagram}
\end{figure}

To begin, we employ screening to determine a set of significant two index
operators that must be considered on each block, according to the following
criterion: 
\begin{align}
A_{ij(\in \text{blk})}^{\text{blk}}& \text{discard if $v_{ijkl}<$ thresh$_{1}
$ for \textbf{all} $kl\in \text{blk}$}  \notag \\
B_{ij(\in \text{blk})}^{\text{blk}}& \text{discard if $x_{ijkl}<$ thresh$_{1}
$ for \textbf{all} $kl\in \text{blk}$}  \notag \\
P_{ij(\notin \text{blk})}^{\text{blk}}& \text{discard if $v_{ijkl}<$ thresh$%
_{1}$ for \textbf{all} $kl\in \text{blk}$}  \notag \\
Q_{ij(\notin \text{blk})}^{\text{blk}}& \text{discard if $x_{ijkl}<$ thresh$%
_{1}$ for \textbf{all} $kl\in \text{blk.}$}  \label{screeneqn}
\end{align}%
In a DMRG block configuration, there are four kinds of blocks: the left
block $L$, an orbital to be blocked with the left block $\circ _{L}$, an
orbital to be blocked with the right block $\circ _{R}$, and the right block 
$R$ (Fig. \ref{fig:blockdiagram}). Without screening, the number of
two-index operators that must be considered on each block is $O(k^{2})$, but
in each case this is reduced to $O(k)$ after screening since eqns. (\ref%
{screeneqn}) require centres $i,j$ to be close in space. Since the number of
operators is reduced, we also reduce the memory cost to $O(M^{2}k)$ per
sweep iteration (block configuration). The disk usage is reduced to $%
O(M^{2}k^{2})$ per sweep.

Next, we consider the computational costs of the different manipulations
involving the two index operators in each of the three stages of a sweep
iteration: (1) blocking, (2), solving for the wavefunction, and (3)
decimation:

\begin{enumerate}
\item {\textit{Blocking}:} Here we construct representations of the
operators in the tensor product space of a large block and an additional
orbital; for concreteness, we take the large block as the left block $L$,
and the additional orbital as $\circ _{L}$, and we consider the operator $%
P_{ij}$. First, we accumulate $P_{ij}^{L},P_{ij}^{\circ _{L}}$ in the new
space $\{L\}\otimes \{\circ _{L}\}$; for each such term, the accumulation
requires $O(M^{2})$ time. Since there are $O(k)$ screened $P_{ij}$ operators
on both blocks $L$ and $\circ _{L}$, in total this requires $O(M^{2}k)$ time
per blocking step and thus $O(M^{2}k^{2})$ time per sweep. Next, we sum over
the new terms appearing in eqns. (\ref{compop1}), (\ref{compop2}) that arise
from the combinations $k\in L,l\in \circ _{L}$ and $k\in \circ _{L},l\in L$;
each such term requires $O(M^{2})$ time per blocking step. Without screening
the number of new terms become $O(k)$, but with screening we discard any
contributions where $v_{ijkl}<\text{thresh}_{2}\text{ for \textbf{all} }kl%
\text{ $\in $ \text{blk}}$ and this decreases the number of new terms to $%
O(1)$ for each significant $P_{ij}$ operator per blocking step.
Consequently, the time to accumulate the additional contributions is $%
O(M^{2})\times O(1)\times \text{no. significant $P_{ij}$ }=O(M^{2}k)$ per
blocking step, or $O(M^{2}k^{2})$ time per sweep. Repeating this analysis
for the $A_{ij},B_{ij},Q_{ij}$ operators, we observe that these also involve 
$O(M^{2}k^{2})$ time per sweep.

\item {\textit{Solving for the wavefunction}:} In an iterative Davidson
algorithm \cite{davidson}, the contributions of $P_{ij}, Q_{ij}$ to the
Hamiltonian matrix multiply takes the form $\sum_{ij} (P_{ij}^{L\circ_L}
\otimes {A^\dag_{ij}}^{\circ_R R}) |\Psi\rangle, \sum_{ij}
(Q_{ij}^{L\circ_L} \otimes B_{ij}^{\circ_R R}) |\Psi\rangle$. Each $\otimes$
requires $O(M^3)$ time, and thus the overall cost is determined by the
number of $ij$ indices to sum over. From the screening criterion $\text{%
thresh}_1$, this is $O(k)$ for each block configuration, and thus the total
time for a single Hamiltonian multiply takes $O(M^3 k)$, or $O(M^3 k^2)$ per
sweep.

\item {\textit{Decimation}:} In the decimation for each two index operator,
each transformation takes $O(M^{3})$ time. After screening, only $O(k)$ $ij$
indices need be considered per block, and thus the time to transform all $%
A_{ij},B_{ij},P_{ij},Q_{ij}$ operators is $O(M^{3}k)$ per renormalisation
step, or $O(M^{3}k^{2})$ per sweep.
\end{enumerate}

\begin{table*}[tbp]
\caption{Time, memory, and disk costs associated with the two-index
operators in the original DMRG and screened LDMRG algorithms. The two-index
operators determine the asymptotic computational costs of the algorithm.}%
\begin{tabular}{c|cc|cc|cc|cc|cc}
\hline\hline
Operator & \multicolumn{2}{c}{Blocking} & \multicolumn{2}{c}{Solving} & 
\multicolumn{2}{c}{Decimation} & \multicolumn{2}{c}{Memory} & 
\multicolumn{2}{c}{Disk} \\ \hline
& DMRG & LDMRG & DMRG & LDMRG & DMRG & LDMRG & DMRG & LDMRG & DMRG & LDMRG
\\ \hline
$A_{ij},B_{ij}$ & $M^{2}k^{3}$ & $M^{2}k^{2}$ & $M^{3}k^{3}$ & $M^{2}k^{2}$
& $M^{3}k^{3}$ & $M^{3}k^{2}$ & $M^{2}k^{2}$ & $M^{2}k$ & $M^{2}k^{3}$ & $%
M^{2}k^{2}$ \\ 
$P_{ij},Q_{ij}$ & $M^{2}k^{4}$ & $M^{2}k^{2}$ & $M^{3}k^{3}$ & $M^{2}k^{2}$
& $M^{3}k^{3}$ & $M^{3}k^{2}$ & $M^{2}k^{2}$ & $M^{2}k$ & $M^{2}k^{3}$ & $%
M^{2}k^{2}$ \\ \hline\hline
\end{tabular}%
\label{tab:screentab}
\end{table*}

In summary, integral screening in the LDMRG reduces
the total computation cost per sweep to $O(M^{3}k^{2})+O(M^{2}k^{2})$ time
(i.e. quadratic scaling, since M in long molecules is independent of system
size for a chosen accuracy and hence a constant), $O(M^{2}k)$ memory, and $%
O(M^{2}k^{2})$ disk. Table \ref{tab:screentab} summarises the key operations
and costs of the screened algorithm. 

Finally, we note that the above screening procedure is easily combined with
the parallelised algorithm employed in our previous calculations \cite%
{CHAN:2004:_dmrg}. Once the list of screened $ij$ indices is determined via
eqns. \ref{screeneqn}, the significant operators are distributed over the
processors, and all manipulations involving these operators are then carried
out in parallel. This screened parallelised algorithm has been employed to
perform the calculations described in the current work.

\section{Numerical analysis of the LDMRG in long systems}

\label{sec:calculations}

\begin{figure}[tbp]
\caption{Geometries of chemical systems used in the LDMRG study in sec. 
\protect\ref{sec:calculations}. }\includegraphics[width=8cm]{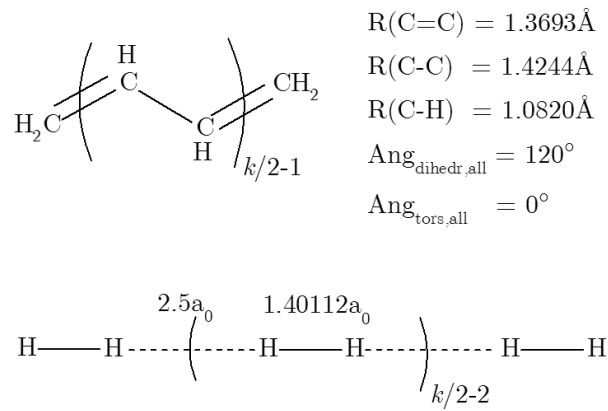}
\label{fig:polyenegeom}
\end{figure}

In the current section, we report our numerical investigations of (i) the
accuracy and extensivity of the LDMRG ansatz in long molecules, (ii)
computational performance of the quadratic-scaling algorithm and robustness
of the screening criteria, (iii) convergence of the LDMRG ansatz, and (iv)
errors compared against standard correlation methods. We have chosen two
classes of systems as representative \textquotedblleft
long\textquotedblright\ molecules: planar all-trans-polyenes $\text{C}_{k}%
\text{H}_{k+2}$ ranging from $k=4,8,\ldots ,48$ (modelled in the $\pi _{z}$%
-active space) and hydrogen molecule chains $(\text{H}_{2})_{k/2}$ ranging
from $k=10,20,\ldots ,100$. The geometries of the polyenes (based on \cite%
{paz}) and hydrogen chains are given in fig. \ref{fig:polyenegeom}. We note
that although the bond-lengths in the hydrogen molecule chains are
alternating, the molecules are still spaced sufficiently closely to be
interacting. 

\subsection{Computational details}

\label{sec:compdetails}

All electronic integrals were obtained using the \textsc{Psi3.2} package 
\cite{PSI3}. We used an STO-3G minimal basis for the polyene calculations
and an STO-6G minimal basis for the hydrogen molecular chains \cite{sto,
emsl}. Polyene calculations were performed in the $\pi _{z}$-active space
spanned by one $p_{z}$ orbital on each carbon center, with each carbon atom
contributing one electron; thus in $\text{C}_{k}\text{H}_{k+2}$ we used a $%
(k,k)$ active space. The remaining electrons were placed in doubly occupied
Restricted Hartree-Fock (RHF) orbitals generated by the \textsc{PSI3.2}
program. Calculations on the hydrogen chains correlated all electrons.

We used a localized orthonormal basis as input to the LDMRG calculations. In both the polyenes and hydrogen chains this was obtained by symmetrically orthonormalising ($S^{-1/2}$)
the atomic orbital basis. The orthonormalised orbitals were then ordered in
their natural topological order i.e. in the order of their originating atoms
along the chain. Since each atom contributes only one basis function, this
ordering is unique.

The LDMRG calculations were performed with the parallel \textsc{Block} code 
\cite{CHAN:2004:_dmrg} with integral screening as described in sec. \ref%
{sec:quadraticscaling}, on 4-18 processors. Except where stated otherwise
(see sec. \ref{sec:scaling}), we applied screening thresholds of $\text{%
thresh}_{1}=10^{-7}$E$_{\text{h}}$ and $\text{thresh}_{2}=10^{-20}$E$_{\text{%
h}}$. No spatial symmetry was used. DMRG sweeps were performed with
progressively increasing $M$ values (a sweep schedule) and a small amount of
random noise (between $10^{-6}-10^{-9}$ in the matrix norm) was added to the
density matrix in the early sweeps ($M\leqslant 100$) to prevent loss of
quantum numbers \cite{CHAN-HEA:2002:_dmrg,CHAN:2004:_dmrg}. A typical
schedule to obtain $M=50,100,250$ DMRG energies is as follows: sweeps 1-6: $%
M=50$ (with noise), sweeps 7-12: $M=50$, sweeps 13-18: $M=100$ (with noise),
sweeps 19-24: $M=100$, sweeps 25-30: $M=250$. We have converged our LDMRG
energies to 8 significant figures; unconverged digits are denoted in
italics. Because of the complexity of the ab initio DMRG method and the non-linearity of the optimisation, there is a small dependence of the DMRG
energies on the precise computational setup (e.g. the way in which $M$ is
increased in sweeps) which may lead to some variation in the last
significant digit.

\subsection{Accuracy and Extensivity of the DMRG ansatz}

In Tables \ref{tab:poly_en} and \ref{tab:h2_n_en} we present the energies
obtained with our quadratic scaling LDMRG algorithm for the
all-trans-polyene series and hydrogen molecular chains. For comparison, we
also present second-order M\o ller-Plesset (MP2) and Coupled Cluster
calculations (CCSD, CCSD(T)) obtained using the \textsc{Psi3.2} (hydrogen
chains) and \textsc{Dalton} \textsc{2.0 }\cite{dalton} (active-space
polyenes) packages.

In the largest $M$ LDMRG calculations shown, the correlation energies are
exact correlation energies for the many-particle Schr\"{o}dinger equation to
the digits displayed. For example, in the polyenes, calculations at the
LDMRG(500) and LDMRG(1000) level did not change the energy in the $\mu $E$_{%
\text{h}}$-range. To confirm the exactness of our LDMRG calculations, we
also performed explicit active space FCI calculations (using \textsc{Molpro
2002.6 } \cite{molpro}) for C$_{4}$H$_{6}$ and $\text{C}_{6}\text{H}_{8}$
and obtained agreement to all displayed digits. Following the discussion in
sec. \ref{sec:compdetails}, the hydrogen molecular chain energies are
presented to 10$\mu $E$_{\text{h}}$ precision, corresponding to eight
significant figures in the electronic energy of the longer chains. There are
only improvements of the order of 1$\mu $E$_{\text{h}}$ when going to
LDMRG(100) and thus the LDMRG(50) correlation energies for the hydrogen
molecular chains are exact to the digits displayed.

The largest Hilbert space considered (for the (H$_{2}$)$_{50}$ system
containing 100 electrons in 100 orbitals) has dim($\mathcal{H}$)=10$^{58}$.
That we are able to obtain a numerically exact correlation energy with the
LDMRG illustrates the compactness of the LDMRG description in systems that
are still interacting but have finite correlation lengths, which allows us
to keep $M$ fixed as the system size grows (see sec. \ref{sec:mps}). A
related feature of the LDMRG ansatz is that of size-consistency/extensivity
of the energy, which we now discuss. 
\begin{figure}[tbp]
\caption{All-trans-polyenes:\ Active space correlation energy from MP2,
CCSD, CCSD(T) and LDMRG(250) as a function of polyene chain length. On the
scale of the graph, the LDMRG and CC results nearly overlap. }
\label{fig:11_poly}\includegraphics[width=9cm]{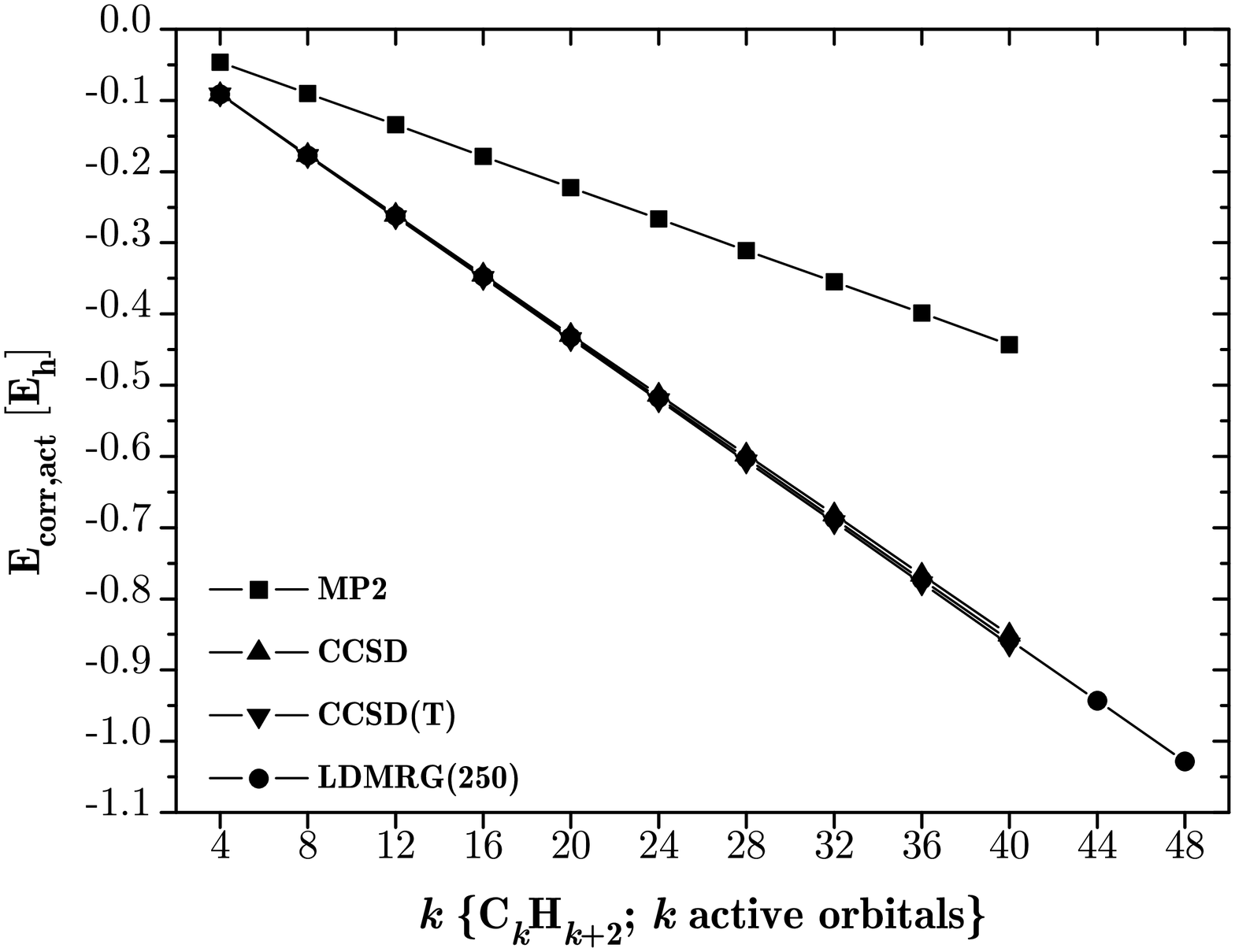}
\end{figure}
\begin{figure}[tbp]
\caption{All-trans-polyenes: Exact (LDMRG(250)) total energy per
additionally introduced C$_{4}$H$_{4}$-unit. The inlay shows the active
space correlation energy at CCSD, CCSD(T) and LDMRG(250) level of theory per
additionally introduced C$_{4}$H$_{4}$-unit.}
\label{fig:14_poly}\centerline{\includegraphics[width=9cm]{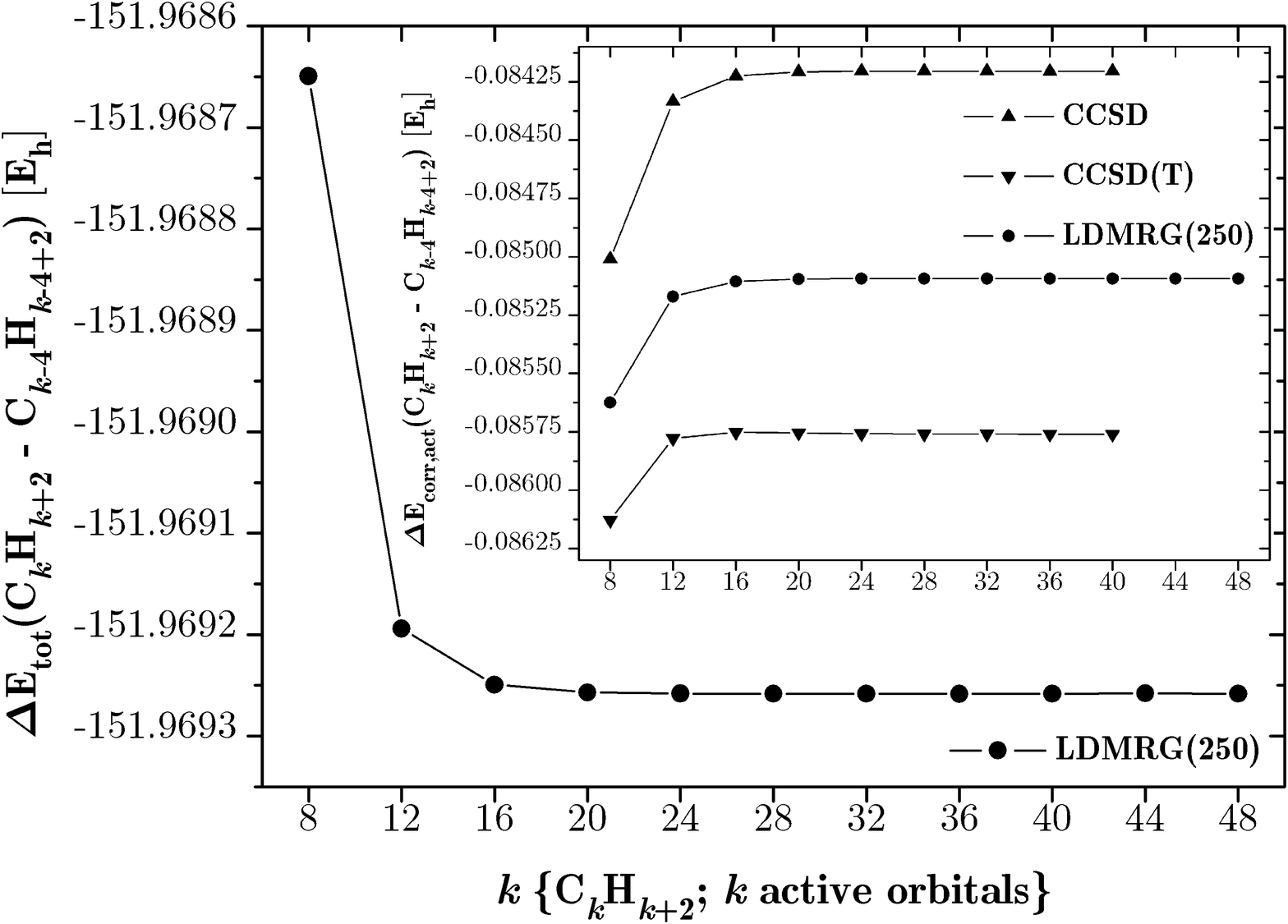}}
\end{figure}

In fig. \ref{fig:11_poly}, we plot the active space correlation energy E$_{%
\text{corr,act}}$ as a function of polyene chain length. A clear linear
relationship between chain length and correlation energy is observed. Fig. %
\ref{fig:14_poly} shows in detail how the active space correlation energy as
well as the total energy E$_{\text{tot}}$ per additionally introduced C$_{4}$%
H$_{4}$-unit converges to a constant in the limit of long polyenes. 
\begin{figure}[tbp]
\caption{All-trans-polyenes: Relative errors in the active space correlation
energies for LDMRG with various $M$ (compared to the exact \textit{\ }%
LDMRG(250) results). The black marked curves are the errors of MP2, CCSD,
and CCSD(T) as reference. The plot shows absolute magnitudes in logarithmic
scale. }
\label{fig:17_poly}\centerline{\includegraphics[width=9cm]{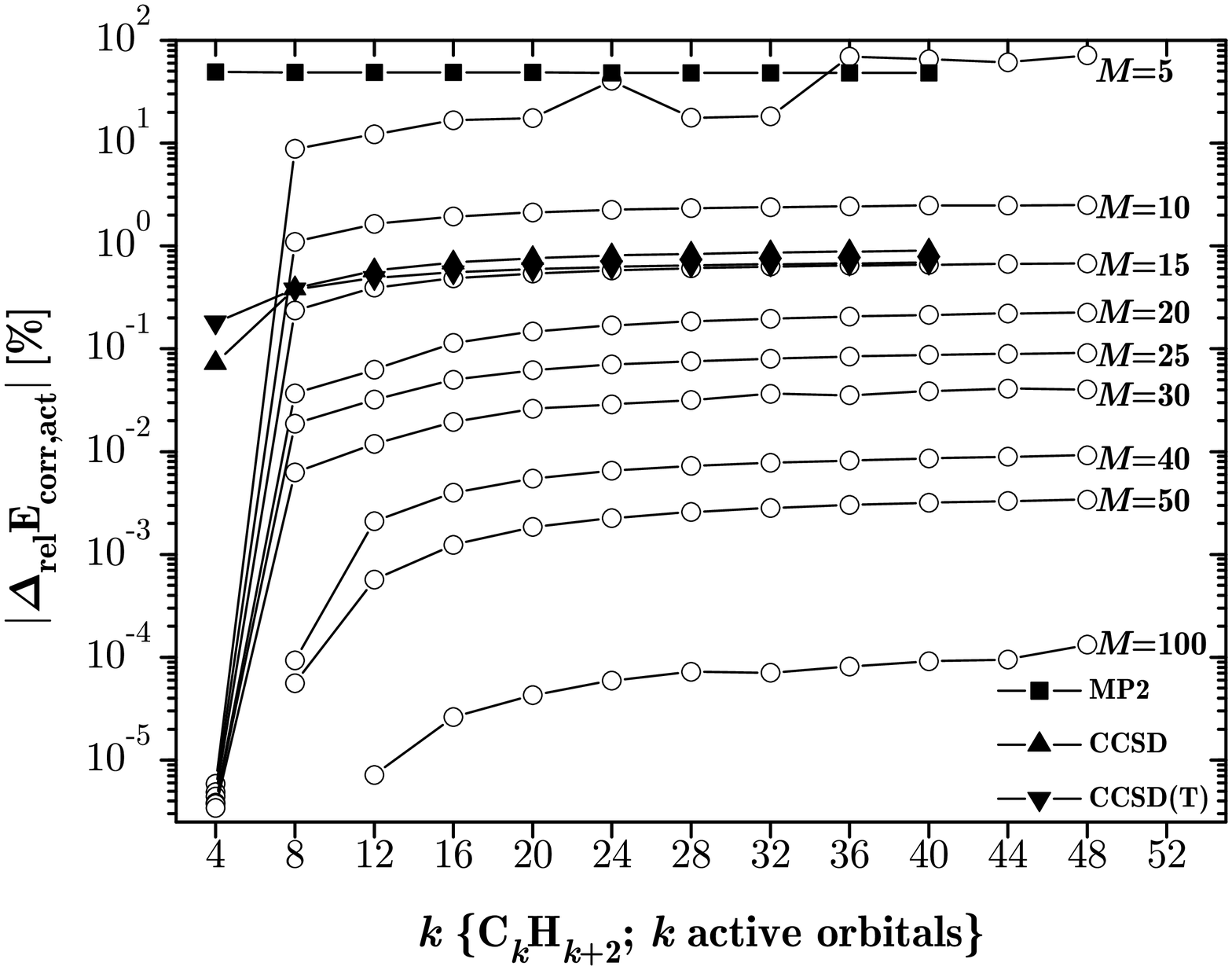}}
\end{figure}

We also performed a series of lower accuracy LDMRG calculations for the
polyenes, with $M=5-40$ states. Due to the variational nature of the DMRG,
these approach the exact energy from above. Fig. \ref{fig:17_poly} presents
the logarithm of the percentage error in the correlation energy relative to
the "exact" LDMRG(250) energies, as a function of chain length. In small
systems the LDMRG calculations are exact, since the DMRG states span the
whole $N$-particle space. In the longer polyenes, the percentage errors
increase to a saturating value, demonstrating the size-extensivity of the
approximate LDMRG calculations. Similar observations can be made for the
hydrogen molecular chains.

\begin{table*}[tbp] \centering%
\caption{All-trans-polyenes: Dimension of the FCI determinant space, total
RHF energy, RHF active space electronic energy; active space correlation
energies at MP2, CCSD, CCSD(T) and different LDMRG levels of theory. All
energies are given in hartrees.}\label{tab:poly_en}%
\begin{tabular}{clcccccccc}
\hline\hline
&  &  &  & E$_{\text{corr,act}}$ &  &  &  &  &  \\ \cline{5-10}
Molecule & dim($\mathcal{H}$)\footnote{%
dim($\mathcal{H}$)$=\binom{k_{\alpha }}{N_{\alpha }}\binom{k_{\beta }}{%
N_{\beta }}$, i.e. here dim($\mathcal{H}$)$=\binom{k/2}{k/2}^{2}$. No point
group symmetry used.} & E$_{\text{RHF}}$ & E$_{\text{RHF,el,act}}$\footnote{%
The active space electronic energy contains the core-active Coulomb and
exchange contributions, but no nuclear repulsion. } & MP2 & CCSD & CCSD(T) & 
LDMRG(50) & LDMRG(100) & LDMRG(250)\footnote{%
All calculations with $M\geqslant 250$ converged.} \\ \hline
C$_{4}$H$_{6}$ & 3.6$\times $10$^{1}$ & \multicolumn{1}{r}{-153.006 364} & 
\multicolumn{1}{r}{-3.169 490} & \multicolumn{1}{r}{-0.046 529} & 
\multicolumn{1}{r}{-0.091 435} & \multicolumn{1}{r}{-0.091 668} & 
\multicolumn{1}{r}{-0.091 502} & \textit{conv.}\footnote{%
"\textit{conv.}" denotes converged results, where increased $M$\ did not
change the significant figures in the energy. } & \textit{conv.} \\ 
C$_{8}$H$_{10}$ & 4.9$\times $10$^{3}$ & \multicolumn{1}{r}{-304.889 389} & 
\multicolumn{1}{r}{-8.426 391} & \multicolumn{1}{r}{-0.090 346} & 
\multicolumn{1}{r}{-0.176 445} & \multicolumn{1}{r}{-0.177 797} & 
\multicolumn{1}{r}{-0.177 127} & \textit{conv.} & \textit{conv.} \\ 
C$_{12}$H$_{14}$ & 8.5$\times $10$^{5}$ & \multicolumn{1}{r}{-456.773 412} & 
\multicolumn{1}{r}{-14.589 838} & \multicolumn{1}{r}{-0.134 320} & 
\multicolumn{1}{r}{-0.260 779} & \multicolumn{1}{r}{-0.263 575} & 
\multicolumn{1}{r}{-0.262 296} & \multicolumn{1}{r}{-0.262 297} & \textit{%
conv.} \\ 
C$_{16}$H$_{18}$ & 1.7$\times $10$^{8}$ & \multicolumn{1}{r}{-608.657 556} & 
\multicolumn{1}{r}{-21.345 452} & \multicolumn{1}{r}{-0.178 366} & 
\multicolumn{1}{r}{-0.345 003} & \multicolumn{1}{r}{-0.349 327} & 
\multicolumn{1}{r}{-0.347 399} & \multicolumn{1}{r}{-0.347 403} & \textit{%
conv.} \\ 
C$_{20}$H$_{22}$ & 3.4$\times $10$^{10}$ & \multicolumn{1}{r}{-760.541 718}
& \multicolumn{1}{r}{-28.542 181} & \multicolumn{1}{r}{-0.222 434} & 
\multicolumn{1}{r}{-0.429 210} & \multicolumn{1}{r}{-0.435 082} & 
\multicolumn{1}{r}{-0.432 490} & \multicolumn{1}{r}{-0.432 498} & \textit{%
conv.} \\ 
C$_{24}$H$_{26}$ & 7.3$\times $10$^{12}$ & \multicolumn{1}{r}{-912.425 883}
& \multicolumn{1}{r}{-36.090 721} & \multicolumn{1}{r}{-0.266 507} & 
\multicolumn{1}{r}{-0.513 414} & \multicolumn{1}{r}{-0.520 840} & 
\multicolumn{1}{r}{-0.517 579} & \multicolumn{1}{r}{-0.517 591} & \textit{%
conv.} \\ 
C$_{28}$H$_{30}$ & 1.6$\times $10$^{15}$ & \multicolumn{1}{r}{-1064.310 048}
& \multicolumn{1}{r}{-43.931 953} & \multicolumn{1}{r}{-0.310 582} & 
\multicolumn{1}{r}{-0.597 618} & \multicolumn{1}{r}{-0.606 599} & 
\multicolumn{1}{r}{-0.602 668} & \multicolumn{1}{r}{-0.602 684} & \textit{%
conv.} \\ 
C$_{32}$H$_{34}$ & 3.6$\times $10$^{17}$ & \multicolumn{1}{r}{-1216.194 214}
& \multicolumn{1}{r}{-52.023 816} & \multicolumn{1}{r}{-0.354 658} & 
\multicolumn{1}{r}{-0.681 822} & \multicolumn{1}{r}{-0.692 358} & 
\multicolumn{1}{r}{-0.687 757} & \multicolumn{1}{r}{-0.687 777} & \textit{%
conv.} \\ 
C$_{36}$H$_{38}$ & 8.2$\times $10$^{19}$ & \multicolumn{1}{r}{-1368.078 379}
& \multicolumn{1}{r}{-60.334 842} & \multicolumn{1}{r}{-0.398 734} & 
\multicolumn{1}{r}{-0.766 027} & \multicolumn{1}{r}{-0.778 118} & 
\multicolumn{1}{r}{-0.772 846} & \multicolumn{1}{r}{-0.772 870} & \textit{%
conv.} \\ 
C$_{40}$H$_{42}$ & 1.9$\times $10$^{22}$ & \multicolumn{1}{r}{-1519.962 544}
& \multicolumn{1}{r}{-68.840 593} & \multicolumn{1}{r}{-0.442 810} & 
\multicolumn{1}{r}{-0.850 231} & \multicolumn{1}{r}{-0.863 879} & 
\multicolumn{1}{r}{-0.857 935} & \multicolumn{1}{r}{-0.857 962} & 
\multicolumn{1}{r}{-0.857 963} \\ 
C$_{44}$H$_{46}$ & 4.4$\times $10$^{24}$ & \multicolumn{1}{r}{-1671.846 710}
& \multicolumn{1}{r}{-77.521 543} & \footnote{\label{footnote1}C$_{44}$H$%
_{46}$ and C$_{48}$H$_{50}$ could not be computed by \textsc{Dalton} due to
an address limitation. } \  & $^{\text{\ref{footnote1}}}$ & $^{\text{\ref%
{footnote1}}}$ & \multicolumn{1}{r}{-0.943 024} & \multicolumn{1}{r}{-0.943
055} & \multicolumn{1}{r}{-0.943 056} \\ 
C$_{48}$H$_{50}$ & 1.0$\times $10$^{27}$ & \multicolumn{1}{r}{-1823.730 875}
& \multicolumn{1}{r}{-86.361 727} & $^{\text{\ref{footnote1}}}$ & $^{\text{%
\ref{footnote1}}}$ & $^{\text{\ref{footnote1}}}$ & \multicolumn{1}{r}{-1.028
113} & \multicolumn{1}{r}{-1.028 147} & \multicolumn{1}{r}{-1.028 149} \\ 
\hline\hline
\end{tabular}%
\end{table*}%

\begin{table*}[tbp] \centering%
\caption{(H$_{2}$)$_{k/2}$-chains: Dimension of the FCI determinant space, total
RHF energy, RHF electronic energy; correlation energies at MP2, CCSD,
CCSD(T) and LDMRG(50) levels of theory. All energies are given in hartrees.}%
\label{tab:h2_n_en}%
\begin{tabular}{cccccccc}
\hline\hline
&  &  &  & E$_{\text{corr}}$ &  &  &  \\ \cline{5-8}
Molecule & dim($\mathcal{H}$)\footnote{%
No point group symmetry used.} & E$_{\text{RHF}}$ & E$_{\text{RHF,el}}$ & MP2
& CCSD & CCSD(T) & LDMRG(50)\footnote{%
All calculations with $M\geqslant 50$ converged.} \\ \hline
(H$_{2}$)$_{5}$ & 6.4$\times $10$^{4}$ & \multicolumn{1}{r}{-5.553 26} & 
-16.036 48 & \multicolumn{1}{r}{-0.068 34} & \multicolumn{1}{r}{-0.101 93} & 
\multicolumn{1}{r}{-0.102 04} & \multicolumn{1}{r}{-0.102 09} \\ 
(H$_{2}$)$_{10}$ & 3.4$\times $10$^{10}$ & \multicolumn{1}{r}{-11.088 22} & 
-38.784 11 & \multicolumn{1}{r}{-0.137 53} & \multicolumn{1}{r}{-0.203 77} & 
\multicolumn{1}{r}{-0.204 04} & \multicolumn{1}{r}{-0.204 15} \\ 
(H$_{2}$)$_{15}$ & 2.4$\times $10$^{16}$ & \multicolumn{1}{r}{-16.623 18} & 
-64.210 83 & \multicolumn{1}{r}{-0.206 72} & \multicolumn{1}{r}{-0.305 61} & 
\multicolumn{1}{r}{-0.306 03} & \multicolumn{1}{r}{-0.306 21} \\ 
(H$_{2}$)$_{20}$ & 1.9$\times $10$^{22}$ & \multicolumn{1}{r}{-22.158 14} & 
-91.378 72 & \multicolumn{1}{r}{-0.275 91} & \multicolumn{1}{r}{-0.407 44} & 
\multicolumn{1}{r}{-0.408 02} & \multicolumn{1}{r}{-0.408 26} \\ 
(H$_{2}$)$_{25}$ & 1.6$\times $10$^{28}$ & \multicolumn{1}{r}{-27.693 11} & 
-119.841 65 & \multicolumn{1}{r}{-0.345 10} & \multicolumn{1}{r}{-0.509 28}
& \multicolumn{1}{r}{-0.510 01} & \multicolumn{1}{r}{-0.510 32} \\ 
(H$_{2}$)$_{30}$ & 1.4$\times $10$^{34}$ & \multicolumn{1}{r}{-33.228 07} & 
-149.336 70 & \multicolumn{1}{r}{-0.414 29} & \multicolumn{1}{r}{-0.611 12}
& \multicolumn{1}{r}{-0.612 01} & \multicolumn{1}{r}{-0.612 38} \\ 
(H$_{2}$)$_{35}$ & 1.3$\times $10$^{40}$ & \multicolumn{1}{r}{-38.763 03} & 
-179.690 11 & \multicolumn{1}{r}{-0.483 48} & \multicolumn{1}{r}{-0.712 95}
& \multicolumn{1}{r}{-0.714 00} & \multicolumn{1}{r}{-0.714 4\textit{4}%
\footnote{%
This result lies very close on the rounding \ border, LDMRG(50) rounds the
last digit to 4, LDMRG(100) to 5.}} \\ 
(H$_{2}$)$_{40}$ & 1.2$\times $10$^{46}$ & \multicolumn{1}{r}{-44.297 99} & 
-210.778 35 & \multicolumn{1}{r}{-0.552 67} & \multicolumn{1}{r}{-0.814 79}
& \multicolumn{1}{r}{-0.815 99} & \multicolumn{1}{r}{-0.816 49} \\ 
(H$_{2}$)$_{45}$ & 1.1$\times $10$^{52}$ & \multicolumn{1}{r}{-49.832 95} & 
-242.509 08 & \multicolumn{1}{r}{-0.621 87} & \multicolumn{1}{r}{-0.916 63}
& \multicolumn{1}{r}{-0.917 98} & \multicolumn{1}{r}{-0.918 55} \\ 
(H$_{2}$)$_{50}$ & 1.0$\times $10$^{58}$ & \multicolumn{1}{r}{-55.367 92} & 
-274.810 58 & \multicolumn{1}{r}{-0.691 06} & \multicolumn{1}{r}{-1.018 47}
& \multicolumn{1}{r}{-1.019 98} & \multicolumn{1}{r}{-1.020 61} \\ 
\hline\hline
\end{tabular}%
\end{table*}%

\subsection{Computational Scaling and Screening Robustness}

\label{sec:scaling} 
\begin{figure}[tbp]
\caption{(H$_{2}$)$_{k/2}$-chains (circles) and all-trans-polyenes
(squares): Asymptotic timing data (i.e. total time per sweep) of LDMRG with $%
M=50$ (filled marks) and $M=250$ (unfilled marks) for 10$^{-7}$E$_{\text{h}}$%
-screening in log-log-representation with linear fit. }
\label{fig:27_h2n}\centerline{\includegraphics[width=9cm]{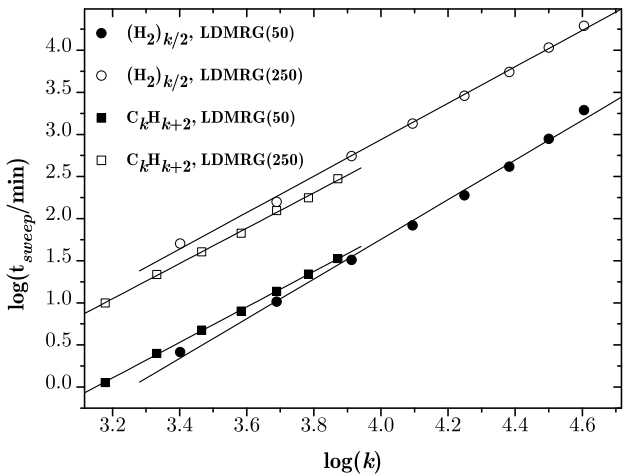}}
\end{figure}

In Figure \ref{fig:27_h2n} we present the asymptotic
computational scaling of the sweep time for the LDMRG\ calculations as a
function of the number of active orbitals for the polyenes and hydrogen
molecular chains. Here sweep times were measured after several sweeps at a
given $M$ level had been performed, to remove the bias that occurs
immediately after a transition from a lower $M$ calculation in the sweep
schedule.

We fitted the timing data to obtain the computational scaling of LDMRG as a
function of the number of active orbitals. The scaling exponents for the
polyenes and the (H$_{2}$)$_{k/2}$-chains with different $M$ and screening
thresholds are given in tab. \ref{scaletable}.

In the polyenes we find a reduced scaling of near-quadratic order, with an
exponent between $2.1-2.2$. For reasonable screening thresholds (i.e. thresh$%
_{1}$ 10$^{-6}$E$_{\text{h}}-$10$^{-8}$E$_{\text{h}}$) no significant
differences in the scaling is observed. We also do not see a significant
scaling dependence on $M$. In the hydrogen chains a similar reduced scaling
was found, in this case with exponents ranging from $2.2-2.4$. In both
cases, it is clear that the screened LDMRG algorithm has reduced the
computational scaling of the DMRG to quadratic order. As an example of
absolute times per sweep, for the largest system (H$_{2}$)$_{50}$ using 18
2.0 GHz Opteron processors, we required 27min for $M=50$, 37min for $M=100$,
and 73min for $M=250$.

Since the LDMRG employs screening, we should assess the robustness of the
criterion that is used. To this end, we studied the polyene correlation
energies computed with screening thresholds ($\text{thresh}_{1}$) of $%
10^{-6} $E$_{\text{h}}$, $10^{-7}$E$_{\text{h}}$, 10$^{-8}$E$_{\text{h}}$
and 10$^{-20}$E$_{\text{h}}$ (the energy of the latter can be considered
unscreened). A selection of results is presented in tab. \ref%
{tab:screentable}. We observe the correlation energy to be converged at the $%
\mu \text{E}_{\text{h}}$ level with respect to the screening threshold when $%
\text{thresh}_{1}=10^{-7}$E$_{\text{h}}$, which is the reason for using this
setting during this study. In practice, this threshold could be relaxed
for lower accuracy calculations.

\begin{table}[tbp] \centering%
\caption{    All-trans-polyenes: Active space correlation 
energies from LDMRG(250) with screening 
thresholds $10^{-6}$E$_{\text{h}}$ and $10^{-7}$E$_{\text{h}}$; absolute and 
relative errors of $10^{-6}$E$_{\text{h}}$-screening (compared to the exact 
results from $10^{-7}$E$_{\text{h}}$-screening). All energies are given in hartrees.
        }\label{tab:screentable}%
\begin{tabular}{ccccc}
\hline\hline
& E$_{\text{corr,act}}$ &  &  &  \\ \cline{2-5}
Molecule & $($10$^{-6}$E$_{\text{h}})$ & $($10$^{-7}$E$_{\text{h}})$ & $%
\Delta _{\text{abs}}$ & $\Delta _{\text{rel}}$ [\%] \\ \hline
C$_{8}$H$_{10}$ & \multicolumn{1}{r}{-0.177 127} & \multicolumn{1}{r}{%
\textit{conv.}} & 0 & 0 \\ 
C$_{16}$H$_{18}$ & \multicolumn{1}{r}{-0.347 405} & \multicolumn{1}{r}{
-0.347 404} & -0.000 001 & -0.4\textit{6}$\times $10$^{-5}$ \\ 
C$_{32}$H$_{34}$ & \multicolumn{1}{r}{-0.687 765} & \multicolumn{1}{r}{
-0.687 777} & 0.000 008 & 1.5\textit{2}$\times $10$^{-5}$ \\ 
C$_{40}$H$_{42}$ & \multicolumn{1}{r}{-0.857 942} & \multicolumn{1}{r}{
-0.857 963} & 0.000 021 & 3.0\textit{1}$\times $10$^{-5}$ \\ 
C$_{48}$H$_{50}$ & \multicolumn{1}{r}{-1.028 093} & \multicolumn{1}{r}{
-1.028 149} & 0.000 056 & 6.4\textit{1}$\times $10$^{-5}$ \\ \hline\hline
\end{tabular}%
\end{table}%

\begin{table*}[tbp] \centering%
\caption{Asymptotic scaling exponents (with standard error) of LDMRG
depending on $M$ and the screening threshold.}\label{scaletable}

\begin{tabular}{cccc}
\hline\hline
& Scaling exponent &  &  \\ \cline{2-4}
System (thresh$_{1}$ [E$_{\text{h}}$]) & LDMRG(50) & LDMRG(100) & LDMRG(250)
\\ \hline
C$_{k}$H$_{k+2}$ (10$^{-6}$) & \multicolumn{1}{r}{2.12$\pm $0.02} & 
\multicolumn{1}{r}{2.11$\pm $0.01} & \multicolumn{1}{r}{2.07$\pm $0.09} \\ 
C$_{k}$H$_{k+2}$ (10$^{-7}$) & \multicolumn{1}{r}{2.11$\pm $0.02} & 
\multicolumn{1}{r}{2.11$\pm $0.01} & \multicolumn{1}{r}{2.10$\pm $0.03} \\ 
C$_{k}$H$_{k+2}$ (10$^{-8}$) & \multicolumn{1}{r}{2.12$\pm $0.01} & 
\multicolumn{1}{r}{2.07$\pm $0.02} & \multicolumn{1}{r}{2.09$\pm $0.03} \\ 
C$_{k}$H$_{k+2}$ (10$^{-20}$) & \multicolumn{1}{r}{3.27$\pm $0.08} & 
\multicolumn{1}{r}{3.33$\pm $0.10} & \multicolumn{1}{r}{3.53$\pm $0.06} \\ 
(H$_{2}$)$_{k/2}$ (10$^{-7}$) & \multicolumn{1}{r}{2.36$\pm $0.06} & 
\multicolumn{1}{r}{2.18$\pm $0.05} & \multicolumn{1}{r}{2.16$\pm $0.04} \\ 
\hline\hline
\end{tabular}
\end{table*}%

\subsection{Sweep and Error Convergence}

There are two types of convergence in DMRG calculations. The first is the
convergence of the energy as a function of the number of sweeps, holding the
number of DMRG states $M$ fixed. We observed that on average, convergence
was achieved in only $4-6$ sweeps for small $M$ and $2-4$ sweeps for large $%
M $ values (not inclusive of the noise sweeps and the preceding sweeps in
the schedule).

\begin{figure}[tbp]
\caption{All-trans-polyenes:\ Convergence of the relative errors in the
active space correlation energies for LDMRG as a function of $M$ (compared
to the exact LDMRG(250) results). The main plot shows magnitudes in
logarithmic scale over $\log \left( M\right) ^{2}$ (with linear fit), the
inlay shows them over $\log (M)$. }\centerline{%
\includegraphics[width=9cm]{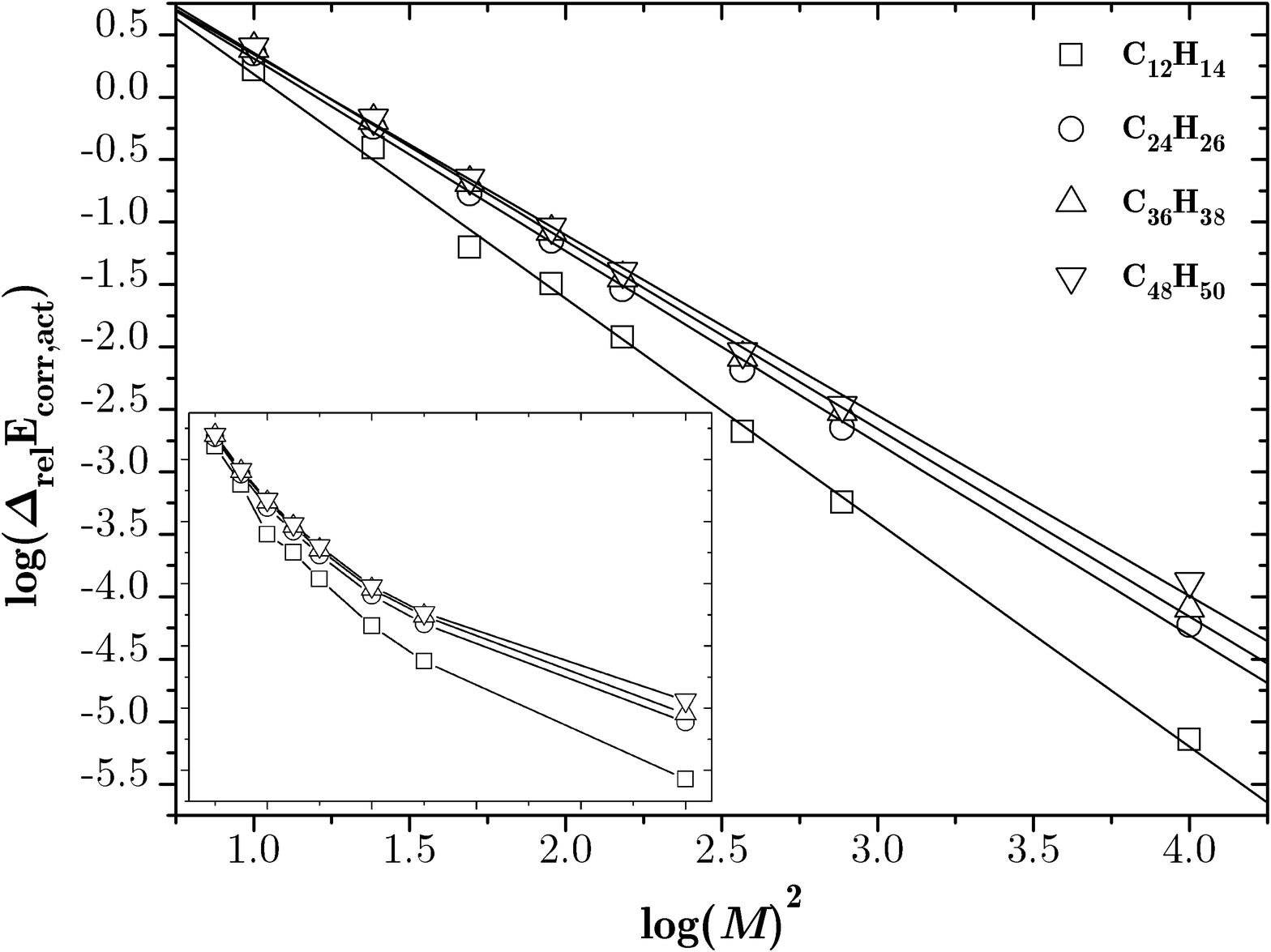}}
\label{fig:18_poly}
\end{figure}

The second type of convergence relates to the approach of the DMRG energy to
the exact energy as the number of retained states $M$ is increased. Here, we
analyse our data for the DMRG calculations on polyenes with different $M$
values. The precise analytic form of the DMRG energy convergence as a
function of $M$ has been a matter of debate in the literature \cite%
{MLPF:2003:_dmrg, CHAN-HEA:2002:_dmrg, chan:2002b, okunishi, reiher}. We
have previously found good agreement with the proposed form of Okunishi et
al. \cite{chan:2002b, okunishi}, 
\begin{equation}
\Delta \text{E}(M)\sim \exp \left[ -\kappa \left( \log M\right) ^{\alpha }%
\right] ,\text{ }\alpha =2  \label{log_formula}
\end{equation}%
which is slower than exponential but still faster than algebraic. In fig. %
\ref{fig:18_poly} we plot the logarithm of the percentage error in the
correlation energy against $(\log M)^{2}$, which shows a clear linear fit.
By contrast, the inlay plots the logarithm of the percentage error against $M$, which demonstrates that the error indeed does not decay
exponentially. Fitting our data (omitting $M=5$) to the general form of eqn. %
\ref{log_formula} we obtained an exponent of $\alpha \sim 1.6-1.8$. Fixing $%
\alpha =2$, we obtain values between $\kappa =1.80\pm 0.03$ (for C$_{12}$H$%
_{14}$) and $\kappa =1.45\pm 0.03$ (for C$_{48}$H$_{50}$).

Corresponding to the rapid energy convergence we also observed a rapid
decrease of the truncated weight of the density matrix as $M$ is increased.
This shows that the local representation is well suited to the chemical system and physical problem at hand \cite{legeza,
LEG-ROD:2003:_dmrg2, SCHOLLWOCK:2005:_dmrg}.

\subsection{Comparison with Perturbation and Coupled Cluster Theories}

Since with our larger $M$ LDMRG calculations we have exact energies at our
disposal we can analyze the errors at the various levels of theory in
more detail.

In the polyene calculations (tab. \ref{tab:poly_en}) the largest DMRG
absolute error is 35$\mu $E$_{\text{h}}$ for the C$_{48}$H$_{50}$ molecule
at the $M=50$ level. This corresponds to $\sim 10^{-3}\%$ of the exact
active space correlation energy, and $\sim 10^{-5}\%$ of the exact total
active space electronic energy. Compared to the Coupled Cluster errors,
LDMRG(50) is already better by 2-3 orders of magnitude. The LDMRG(100) gives
a further order of magnitude improvement, and is essentially exact. In our
more approximate calculations (fig. \ref{fig:17_poly}) we find that LDMRG
with $M=10$ performs better than MP2, and with $M=15$ better than CCSD and
CCSD(T). The results for $M=5$ are not reliable due to loss of important
quantum numbers. 

Surprisingly we observe that the CCSD(T) results lie below the exact
energies computed with LDMRG. Although CCSD(T) is not variational in
general, it is still uncommon to obtain an energy below the exact energy
at an equilibrium geometry. In general the triples correction performed
relatively badly for the polyenes. 

In case of the hydrogen chains, the convergence of the LDMRG with $M$ was
more rapid and results were already exact with $M=50$. CCSD(T) also
performed better in this system, the triples correction improved on CCSD by
1/2 order of magnitude, and the resulting energies were consistently above
the exact energies.

\section{The Metal-Insulator Transition in Linear Hydrogen}

\label{sec:mi}

As an example of a challenging electronic problem, we studied the symmetric
and asymmetric bond stretching in a linear H$_{50}$-chain. In both these
cases, the system transitions from a state with metallic correlations at
compressed geometries to an insulating state with strong multireference
correlation in the dissociation region. This bond breaking process hence
exhibits a varying nature of chemical bonding and electron correlation.

In case of the symmetric dissociation we begin with a uniform bond-distance
between all H-atoms of R=1.0a$_{0}$, and stretch all 49 bonds symmetrically
and simultaneously to R=1.2, 1.4, ..., 4.2a$_{0}$. The final structure
consists of 50 equidistant, nearly-independent H-atoms on a line.

In case of the asymmetric dissociation we distinguish alternating bonds as
intermolecular and intramolecular with R$_{\text{inter}}$ and\ R$_{\text{%
intra}}$. The first geometry is R$_{\text{intra}}$=R$_{\text{inter}}$ =1.4a$%
_{0}$. In the following geometries R$_{\text{intra}}$ is kept fixed at 1.4a$%
_{0}$ while R$_{\text{inter}}$ grows to R$_{\text{inter}}$=1.6, 1.8, ...,
4.2a$_{0}$. The final structure consists of 25 equidistant,
nearly-independent H$_{2}$-molecules at equilibrium bond distance on a line.

We computed the electronic energy using the LDMRG (with up to 1000 states)
in the minimal STO-6G basis, where we correlated all 50 electrons in 50
orbitals.

All calculations were carried out in the STO-6G basis correlating all electrons (50 electrons in 50 orbitals). The LDMRG calculations again used the $S^{-1/2}$ basis.

\subsection{Symmetric Dissociation}

\begin{table*}[tbp] \centering%
\caption{Symmetric dissociation of H$_{50}$: Total RHF
energy, RHF electronic energy; correlation energies at MP2, CCSD, CCSD(T)
and various LDMRG levels of theory. All energies are given in hartrees.}%
\label{tab:sym_en_tab}%
\begin{tabular}{cccccccccc}
\hline\hline
&  &  & E$_{\text{corr}}$ &  &  &  &  &  &  \\ \cline{4-10}
R [a$_{0}$] & E$_{\text{RHF}}$ & E$_{\text{RHF,el}}$ & MP2 & CCSD & CCSD(T)
& LDMRG(50) & LDMRG(100) & LDMRG(250) & LDMRG(500)\footnote{%
All calculations with $M\geqslant 500$ converged.} \\ \hline
1.0 & \multicolumn{1}{r}{-16.864 88} & \multicolumn{1}{r}{-191.825 14} & 
\multicolumn{1}{r}{-0.361 45} & \multicolumn{1}{r}{-0.407 29} & 
\multicolumn{1}{r}{-0.417 39} & \multicolumn{1}{r}{-0.402 \textit{72}} & 
-0.417 27 & -0.419 14 & -0.419 19 \\ 
1.2 & \multicolumn{1}{r}{-22.461 27} & \multicolumn{1}{r}{-168.261 49} & 
\multicolumn{1}{r}{-0.401 83} & \multicolumn{1}{r}{-0.470 11} & 
\multicolumn{1}{r}{-0.483 30} & \multicolumn{1}{r}{-0.475 9\textit{0}} & 
-0.485 21 & -0.486 35 & -0.486 38 \\ 
1.4 & \multicolumn{1}{r}{-25.029 76} & \multicolumn{1}{r}{-150.001 38} & 
\multicolumn{1}{r}{-0.444 73} & \multicolumn{1}{r}{-0.543 03} & 
\multicolumn{1}{r}{-0.559 36} & \multicolumn{1}{r}{-0.557 1\textit{6}} & 
-0.563 30 & -0.564 00 & -0.564 02 \\ 
1.6 & \multicolumn{1}{r}{-26.062 25} & \multicolumn{1}{r}{-135.412 42} & 
\multicolumn{1}{r}{-0.491 88} & \multicolumn{1}{r}{-0.631 18} & 
\multicolumn{1}{r}{-0.650 89} & \multicolumn{1}{r}{-0.652 7\textit{2}} & 
-0.656 74 & -0.657 18 & -0.657 19 \\ 
1.8 & \multicolumn{1}{r}{-26.265 98} & \multicolumn{1}{r}{-123.466 13} & 
\multicolumn{1}{r}{-0.545 50} & \multicolumn{1}{r}{-0.741 67} & 
\multicolumn{1}{r}{-0.765 47} & \multicolumn{1}{r}{-0.769 8\textit{2}} & 
-0.772 42 & -0.772 66 & -0.772 67 \\ 
2.0 & \multicolumn{1}{r}{-26.008 20} & \multicolumn{1}{r}{-113.488 34} & 
\multicolumn{1}{r}{-0.607 89} & \multicolumn{1}{r}{-0.883 29} & 
\multicolumn{1}{r}{-0.912 70} & \multicolumn{1}{r}{-0.916 1\textit{1}} & 
-0.917 76 & -0.917 89 & \textit{conv.}\footnote{%
"\textit{conv.}" denotes converged results, where increased $M$\ did not
change the significant figures in the energy. } \\ 
2.4 & \multicolumn{1}{r}{-24.835 76} & \multicolumn{1}{r}{-97.735 87} & 
\multicolumn{1}{r}{-0.768 83} & \multicolumn{1}{r}{\footnote{%
The Coupled Cluster calculations could not be converged. See text. \label%
{footnote_sym_en_tab}}} & \multicolumn{1}{r}{$^{\text{\ref%
{footnote_sym_en_tab}}}$} & \multicolumn{1}{r}{-1.324 16} & -1.324 77 & 
-1.324 81 & \textit{conv.} \\ 
2.8 & \multicolumn{1}{r}{-23.360 81} & \multicolumn{1}{r}{-85.846 62} & 
\multicolumn{1}{r}{-0.995 30} & \multicolumn{1}{r}{$^{\text{\ref%
{footnote_sym_en_tab}}}$} & \multicolumn{1}{r}{$^{\text{\ref%
{footnote_sym_en_tab}}}$} & \multicolumn{1}{r}{-1.913 81} & -1.913 98 & 
-1.913 99 & \textit{conv.} \\ 
3.2 & \multicolumn{1}{r}{-21.896 33} & \multicolumn{1}{r}{-76.571 41} & 
\multicolumn{1}{r}{-1.307 78} & \multicolumn{1}{r}{$^{\text{\ref%
{footnote_sym_en_tab}}}$} & \multicolumn{1}{r}{$^{\text{\ref%
{footnote_sym_en_tab}}}$} & \multicolumn{1}{r}{-2.671 9\textit{0}} & -2.671
95 & \textit{conv.} & \textit{conv.} \\ 
3.6 & \multicolumn{1}{r}{-20.574 29} & \multicolumn{1}{r}{-69.174 36} & 
\multicolumn{1}{r}{-1.723 32} & \multicolumn{1}{r}{$^{\text{\ref%
{footnote_sym_en_tab}}}$} & \multicolumn{1}{r}{$^{\text{\ref%
{footnote_sym_en_tab}}}$} & \multicolumn{1}{r}{-3.528 46} & -3.528 48 & 
\textit{conv.} & \textit{conv.} \\ 
4.2 & \multicolumn{1}{r}{-18.955 95} & \multicolumn{1}{r}{-60.613 15} & 
\multicolumn{1}{r}{-2.558 99} & \multicolumn{1}{r}{$^{\text{\ref%
{footnote_sym_en_tab}}}$} & \multicolumn{1}{r}{$^{\text{\ref%
{footnote_sym_en_tab}}}$} & \multicolumn{1}{r}{-4.793 76} & \textit{conv.} & 
\textit{conv.} & \textit{conv.} \\ \hline\hline
\end{tabular}%
\end{table*}%

\begin{figure}[tbp]
\caption{Symmetric dissociation of H$_{50}$:\ Potential energy curves from
RHF, MP2, and exact LDMRG. On the scale of the graph, the few available CCSD
and CCSD(T) datapoints were indistinguishable from the LDMRG data.}
\label{fig:31_sym}\centerline{\includegraphics[width=9cm]{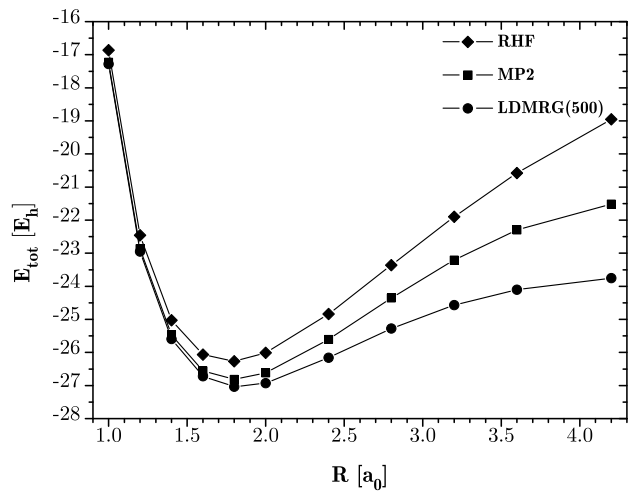}}
\end{figure}

The calculated energies for the symmetric dissociation are summarized in
tab. \ref{tab:sym_en_tab}. The potential energy curves at RHF, MP2, and
exact LDMRG level of theory are plotted in fig. \ref{fig:31_sym}. It can
immediately be seen how the contribution of correlation increases along the
dissociation coordinate: In the dissociation limit the share of the
correlation energy in the total energy grows to $\sim $20\% and in the
electronic energy to $\sim $7\%, which emphasizes the importance of
nondynamic correlation in this problem. 

\begin{figure}[tbp]
\caption{Symmetric dissociation of H$_{50}$: Relative errors in the
correlation energies at MP2, CCSD, CCSD(T),~and different LDMRG levels of
theory (compared to the exact LDMRG results) in logarithmic scale.}%
\centerline{\includegraphics[width=9cm]{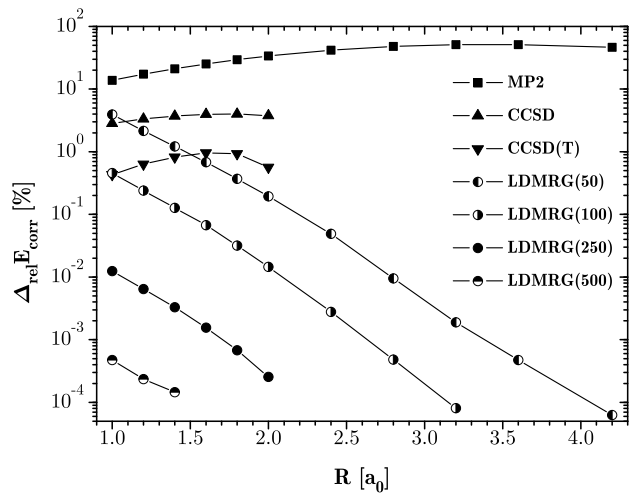}}
\label{fig:35_sym}
\end{figure}

As is expected, RHF and MP2 behave poorly as the chain dissociates. The
Coupled Cluster energies cannot even be converged for bond lengths R$>$2.0a$%
_{0}$. This is a fundamental problem in CC theory that is well documented
e.g. in the work of Takahashi and Paldus and others \cite{paldus1, paldus2,
paldus3} where in one-dimensional systems, even for physically relevant
coupling parameters, the Coupled Cluster doubles equations may have no real
solutions. The correlation energy errors for different methods relative to
the exact LDMRG results are shown in Fig. \ref{fig:35_sym}.

It is understood that we need to retain more states in the LDMRG in the
metallic regime if we start from a local atomic orbital basis, since we need
to capture the delocalisation and long-range off-diagonal correlations \cite%
{SCHOLLWOCK:2005:_dmrg}. We find that both the convergence with the number
of sweeps as well as with $M$ is slower as compared to calculations in the
nonmetallic regime. At R=1.0a$_{0}$ LDMRG(50) is worse than CCSD, LDMRG(100)
slightly worse than CCSD(T), and for R$<$1.6a$_{0}$ LDMRG(50) is still worse
than CCSD(T). In the metallic region LDMRG required $M=500$ to converge to
the numerically exact result. In essence, by using orthonormalised atomic
orbitals, we are starting from a particularly unfavourable one-particle
basis to describe metallic behaviour. By performing the DMRG in a set of
separately localised occupied and virtual orbitals such as Boys orbitals 
\cite{Boys}, we expect that the degradation in efficiency of the DMRG would
be avoided.

\subsection{Asymmetric Dissociation}

\begin{table*}[tbp] \centering%
\caption{Asymmetric dissociation of H$_{50}$: Total RHF energy, RHF electronic
energy; correlation energies at MP2, CCSD, CCSD(T) and various LDMRG levels
of theory. All energies are given in hartrees.}\label{tab:asym_en_tab}%
\begin{tabular}{cccccccccc}
\hline\hline
&  &  & E$_{\text{corr}}$ &  &  &  &  &  &  \\ \cline{4-10}
R$_{\text{inter}}$ [a$_{0}$] & E$_{\text{RHF}}$ & E$_{\text{RHF,el}}$ & MP2
& CCSD & CCSD(T) & LDMRG(50) & LDMRG(100) & LDMRG(250) & LDMRG(500)\footnote{%
All calculations with $M\geqslant 500$ converged.} \\ \hline
1.4 & \multicolumn{1}{r}{-25.029 76} & \multicolumn{1}{r}{-150.001 38} & 
\multicolumn{1}{r}{-0.444 73} & \multicolumn{1}{r}{-0.543 03} & 
\multicolumn{1}{r}{-0.559 36} & \multicolumn{1}{r}{-0.557 1\textit{6}} & 
-0.563 30 & -0.564 00 & -0.564 02 \\ 
1.6 & \multicolumn{1}{r}{-25.963 71} & \multicolumn{1}{r}{-142.811 31} & 
\multicolumn{1}{r}{-0.392 61} & \multicolumn{1}{r}{-0.516 01} & 
\multicolumn{1}{r}{-0.522 30} & \multicolumn{1}{r}{-0.523 02} & -0.523 64 & 
-0.523 67 & \textit{conv.}\footnote{%
"\textit{conv.}" denotes converged results, where increased $M$\ did not
change the significant figures in the energy. } \\ 
1.8 & \multicolumn{1}{r}{-26.617 68} & \multicolumn{1}{r}{-136.573 91} & 
\multicolumn{1}{r}{-0.369 20} & \multicolumn{1}{r}{-0.505 47} & 
\multicolumn{1}{r}{-0.508 73} & \multicolumn{1}{r}{-0.509 38} & -0.509 48 & 
\textit{conv.} & \textit{conv.} \\ 
2.0 & \multicolumn{1}{r}{-27.071 82} & \multicolumn{1}{r}{-131.089 88} & 
\multicolumn{1}{r}{-0.357 01} & \multicolumn{1}{r}{-0.503 04} & 
\multicolumn{1}{r}{-0.504 97} & \multicolumn{1}{r}{-0.505 48} & -0.505 50 & 
\textit{conv.} & \textit{conv.} \\ 
2.4 & \multicolumn{1}{r}{-27.609 24} & \multicolumn{1}{r}{-121.878 99} & 
\multicolumn{1}{r}{-0.346 20} & \multicolumn{1}{r}{-0.507 17} & 
\multicolumn{1}{r}{-0.508 03} & \multicolumn{1}{r}{-0.508 37} & \textit{conv.%
} & \textit{conv.} & \textit{conv.} \\ 
2.8 & \multicolumn{1}{r}{-27.873 62} & \multicolumn{1}{r}{-114.445 23} & 
\multicolumn{1}{r}{-0.341 62} & \multicolumn{1}{r}{-0.512 77} & 
\multicolumn{1}{r}{-0.513 22} & \multicolumn{1}{r}{-0.513 45} & \textit{conv.%
} & \textit{conv.} & \textit{conv.} \\ 
3.2 & \multicolumn{1}{r}{-28.004 68} & \multicolumn{1}{r}{-108.324 99} & 
\multicolumn{1}{r}{-0.338 67} & \multicolumn{1}{r}{-0.516 18} & 
\multicolumn{1}{r}{-0.516 42} & \multicolumn{1}{r}{-0.516 56} & \textit{conv.%
} & \textit{conv.} & \textit{conv.} \\ 
3.6 & \multicolumn{1}{r}{-28.069 65} & \multicolumn{1}{r}{-103.203 54} & 
\multicolumn{1}{r}{-0.336 34} & \multicolumn{1}{r}{-0.517 50} & 
\multicolumn{1}{r}{-0.517 63} & \multicolumn{1}{r}{-0.517 71} & \textit{conv.%
} & \textit{conv.} & \textit{conv.} \\ 
4.2 & \multicolumn{1}{r}{-28.111 00} & \multicolumn{1}{r}{-96.924 54} & 
\multicolumn{1}{r}{-0.333 73} & \multicolumn{1}{r}{-0.517 49} & 
\multicolumn{1}{r}{-0.517 54} & \multicolumn{1}{r}{-0.517 58} & \textit{conv.%
} & \textit{conv.} & \textit{conv.} \\ \hline\hline
\end{tabular}%
\end{table*}%

\begin{figure}[tbp]
\caption{Asymmetric dissociation of H$_{50}$:\ Potential energy curves from
RHF, MP2, and exact LDMRG. On the scale of the graph, the LDMRG, CCSD, and
CCSD(T) curves are indistinguishable. }\centering\centerline{%
\includegraphics*[width=9cm]{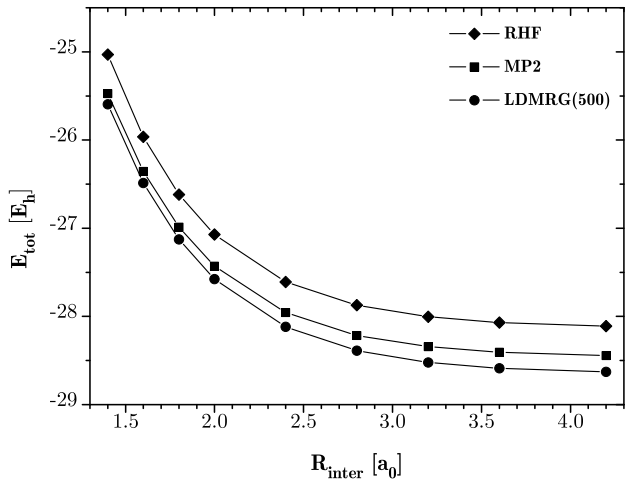}}
\label{fig:41_as}
\end{figure}

\begin{figure}[tbp]
\caption{Asymmetric dissociation of H$_{50}$:\ Relative errors in the
correlation energies at MP2, CCSD, CCSD(T),~and different LDMRG levels of
theory (compared to the exact LDMRG results) in logarithmic scale. }%
\centerline{\includegraphics[width=9cm]{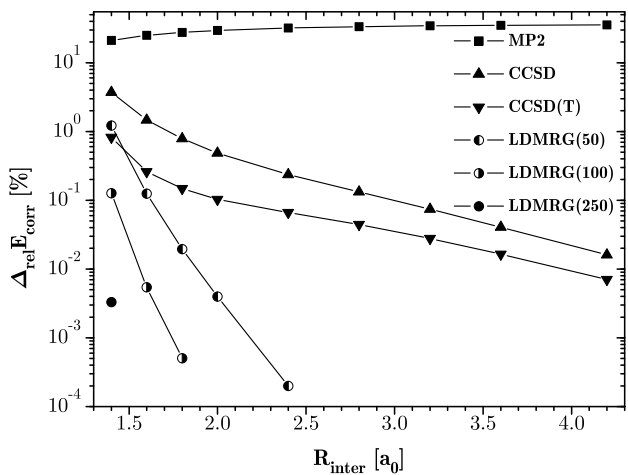}}
\label{fig:45_as}
\end{figure}

The calculated energies for the asymmetric dissociation are summarized in
table \ref{tab:asym_en_tab}. In this system, the restricted Hartree-Fock
reference dissociates correctly (to a set of non-interacting hydrogen
molecules), which can be understood by changing to a localised basis in the
space of restricted occupied orbitals. For this reason, the restricted MP2
and CC theories are also qualitatively correct and we see that their
energies (fig. \ref{fig:41_as}) lie parallel to the exact LDMRG values along
the dissociation curve. Unlike in the symmetric dissociation, the
correlation energy saturates rapidly to $\sim 1.8\%$ of the total energy as
the bonds are stretched. Fig. \ref{fig:45_as} shows how the percentage
errors in the correlation energy for the different methods decrease along
the dissociation coordinate. Again, we see reduced performance of the DMRG
in the metallic regime due to the unsuitability of the underlying orbital
basis, but still a systematic convergence with $M$. For large R$_{\text{inter%
}}$ we observed very rapid convergence with $M$ and number of sweeps, and in
fact for R$_{\text{inter}}$=4.2a$_{0}$ the LDMRG energy was already exact
after 4 noise sweeps with $M=50$. In the limit of complete dissociation, the
CCSD theory becomes exact for this system and this is confirmed by
convergence to the LDMRG results.

\begin{figure}[tbp]
\caption{Asymmetric dissociation of H$_{50}$:\ Half-cross-section of the
LDMRG $\protect\alpha $-one-particle density matrix\ at $\protect\alpha $%
-orbital no. 25. }
\label{fig:46_as}\centerline{\includegraphics[width=9cm]{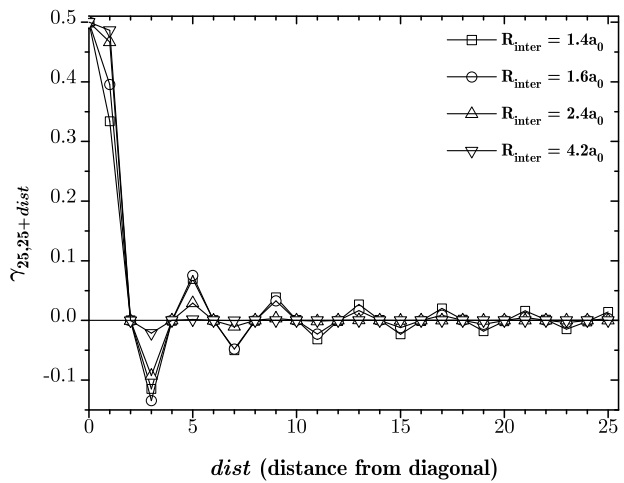}}
\end{figure}

In order to demonstrate the metal-insulator transition more explicitly we
computed the one-particle reduced density matrix $\gamma $ during our LDMRG\
calculations. In fig. \ref{fig:46_as} we have plotted the off-diagonal decay
of the $\alpha $ one-particle density matrix from element $\gamma
_{25,25}\rightarrow \gamma _{25,50}$. In the metallic regime (short R$_{%
\text{inter}}$) we see the long-ranged oscillations in the off-diagonal
elements, while in the insulating regime (long R$_{\text{inter}}$) the
off-diagonal elements decay much more rapidly. A similar picture is obtained
from the density matrix during symmetric dissociation.

\section{Conclusions}

\label{sec:conclusions}

We began this work with the question of how to describe nondynamic
correlation in large systems with the restriction that such systems are
large in only one dimension. In our investigations, we have shown how the
Density Matrix Renormalization Group (DMRG) provides a natural answer to
this problem. The Matrix Product State that underlies the DMRG is a local,
variational, size-consistent/size-extensive, and inherently multireference
ansatz that can efficiently exploit the special structure of
quasi-one-dimensional correlation. Using the intrinsic locality of the
ansatz, we have formulated a \textit{quadratic} scaling DMRG algorithm,
using only a straightforward screening criterion without the imposition of
correlation domains. With this active space method, we could then obtain
numerically exact solutions of the many-particle Schr\"{o}dinger equation
for all-trans-polyenes up to $\text{C}_{48}\text{H}_{50}$ (correlating the $%
\pi _{z}$-electrons) and hydrogen molecular chains up to $(\text{H}_{2})_{50}
$ (correlating 100 electrons in 100 orbitals).

By construction, a unique advantage of the LDMRG as compared to other local
correlation methods is its ability to capture nondynamic correlation. We
can take advantage of locality in multireference problems so long as the
correlation length is finite. We have demonstrated the capability and
efficiency of the LDMRG in these situations by obtaining numerically exact
correlation energies in the metal-to-insulator transition of linear $\text{H}%
_{50}$-chains, where we correlate 50 electrons in 50 orbitals.

With the possibility of accurately capturing nondynamic correlation in long
molecules, we can now begin to address the quantitative description of
strongly interacting states as found in the spectrum of materials such as
the conjugated organic polymers. Here, the natural next step would be to
combine an LDMRG description of the nondynamic correlation in the active $%
\pi $-space with our recent developments in Canonical Transformation Theory 
\cite{CMT}, to incorporate the dynamic correlation that arises in larger
basis sets.

\begin{acknowledgments}
JH is funded by a Kekul\'{e} Fellowship of the Fond der Chemischen Industrie
(Fund of the German Chemical Industry). GKC acknowledges support from
Cornell University and the Cornell Center for Materials Research (CCMR).
Computations were carried out in part on the Nanolab-Cluster of the Cornell
NanoScale Science \& Technology Facility (CNF), supported by NSF ECS
03-05765.
\end{acknowledgments}

\bibliographystyle{aip}
\bibliography{Polyenes}

\end{document}